\documentclass[aps,prd,superscriptaddress,nofootinbib,showpacs,twocolumn]{revtex4}  
\usepackage{amsmath,amssymb}
 
\newcommand{\be}{\begin{equation}} 
\newcommand{\ee}{\end{equation}} 
\newcommand{\bea}{\begin{eqnarray}} 
\newcommand{\eea}{\end{eqnarray}} 
\newcommand{\lb}{\label} 
\newcommand{\bdm}{\begin{displaymath}} 
\newcommand{\edm}{\end{displaymath}} 
\newcommand{\D}{{\rm d}} 
\newcommand{\md}{{\mathrm{d}}} 
\newcommand{\E}{{\rm e}} 
\newcommand{\I}{{\rm i}} 

\def\cF{{\mathcal F}} 
\def\cO{{\mathcal O}} 
 
\begin{document} 
 
\title{Quantum gravitational correction to the Hawking temperature from
 the Lema\^{\i}tre--Tol\-man--Bondi model}

\author{Rabin Banerjee} 
\email{rabin@bose.res.in}
\affiliation{S. N. Bose National Centre for Basic Sciences,
JD Block, Sector III, Salt Lake,
Kolkata-700098, India}
 
 \author{Claus Kiefer}
\email{kiefer@thp.uni-koeln.de}
\affiliation{Institut f\"ur Theoretische Physik,
Universit\"at zu K\"oln, Z\"ulpicher Strasse 77,
50937 K\"oln, Germany}

\author{Bibhas Ranjan Majhi}
\email{bibhas@bose.res.in}
\affiliation{S. N. Bose National Centre for Basic Sciences,
JD Block, Sector III, Salt Lake,
Kolkata-700098, India}
 
\date{\today} 
 
\begin{abstract} 
We solve the quantum constraint equations of the
Lema\^{\i}tre--Tol\-man--Bondi model in a semiclassical
approximation in which an expansion is performed with respect to the
Planck length. We recover in this way the standard expression for the
Hawking temperature as well as its first quantum gravitational
correction. We then interpret this correction in terms of the one-loop
trace anomaly of the energy--momentum tensor and thereby make contact
with earlier work on quantum black holes.
\end{abstract} 
 
\keywords{Hawking temperature \sep canonical quantum gravity \sep
  semiclassical expansion \sep trace anomaly}
\pacs{04.60.Ds, 04.70.Dy}

\maketitle

In the absence of a full quantum theory of gravity, it is of interest
to consider models which could serve as a possible guide in the
construction of such a theory \cite{OUP}. One such model is the 
Lema\^{\i}tre--Tol\-man--Bondi (LTB) model describing the dynamics of a
spherically-symmetric dust cloud \cite{PK}. It has been already used
in a variety of papers dealing with canonical quantization in both the
Wheeler--DeWitt framework and loop quantum cosmology, 
cf. \cite{VWS01,KMHV06,KMHSV07,VTS08,FGK09}
and the references therein. While the full quantization of the LTB
model has not yet been achieved, 
it was at least possible to get insights into the recovery of Hawking radiation
and black-hole entropy from it. 

Our present paper is a continuation of this earlier work.
Our motivation is twofold. First, we want to derive a quantum
gravitational correction to the Hawking temperature through a
semiclassical expansion scheme for the quantum states. Second, we want
to present an interpretation of these correction terms through the ``trace
anomaly'' of the matter energy--momentum tensor, making thereby a
connection to earlier work \cite{CL,DF,BM08,BM09,BM,RMO} in the
context of black holes.   

Let us first present the LTB model.
The spherical gravitational collapse of a dust cloud, in an
asymptotically flat space-time, having energy density
$\epsilon(\tau,\rho)$, is described in comoving coordinates
($\tau,\rho,\theta,\phi$) by the LTB metric, 
\begin{eqnarray}
\D s^2=-\D\tau^2+\frac{(\partial_{\rho}R(\rho,\tau))^2}{1+2E(\rho)}\D\rho^2+
R^2(\rho,\tau)\D\Omega^2 \ .
\label{1.56}
\end{eqnarray}
Inserting this expression into the Einstein field equations leads to,
for vanishing cosmological constant,
\be \label{ltb-eg1} 8\pi G\epsilon(\tau,\rho) = 
\frac{\partial_{\rho}F}{R^2 \partial_{\rho}R} 
\ee 
and 
\be \label{ltb-eg2} 
  (\partial_{\tau}R)^2 = \frac{F}{R} 
  +2E\equiv 1-{\mathcal F}+2E\ , 
\ee 
where $F(\rho)\equiv 2GM(\rho)$, with $M(\rho)$ being the active
gravitational mass within a $\rho=$ constant shell, and
\be\lb{mathcalF} 
{\mathcal F}\equiv 1-\frac{F}{R}\ . 
\ee 
The function 
$E(\rho)$ is the total energy per unit mass within the same shell; 
the marginally bound models are defined by $E(\rho)\equiv 0$. 
The case of collapse is described by $\partial_{\tau}R(\tau,\rho)<0$. 
We set $c=1$ throughout. 

The canonical quantization of the LTB model was developed in \cite{VWS01}  and 
then applied to quantization in a series of papers, see 
\cite{KMHV06,KMHSV07,FGK09}. Although no
full quantization has yet been performed, interesting results have
been obtained at the semiclassical level; they include the recovery of
Hawking radiation plus greybody corrections from solutions to the
Wheeler--DeWitt equation and the momentum constraints (that is, the
quantum constraint equations). Insights into the microscopic
interpretation of black-hole entropy were also obtained, cf.
\cite{entropy} and the references therein. 

The semiclassical approximation scheme is also employed here. We start with the quantum constraint equations \cite{KMHV06},
\bea
l_{\rm P}^4\frac{\delta^2\Psi}{\delta\tau^2}+l_{\rm P}^4\cF
\frac{\delta^2\Psi}{\delta R^2}+\frac{\Gamma^2}{4\cF}\Psi &=& 0
 \ ,\lb{constraint1} \\
\tau'\frac{\delta\Psi}{\delta\tau}+R'\frac{\delta\Psi}{\delta R}
 -\Gamma\left(\frac{\delta\Psi}{\delta\Gamma}\right)^{\prime} &=& 0 \
 ,\lb{constraint2}
\eea
where $\Psi$ is a functional of the dust variable $\tau(r)$ as well as
the gravitational variables $R(r)$ and $\Gamma(r)$, and $l_{\rm
  P}=\sqrt{G\hbar}$ is the Planck length. Here, $r$ is the radial
variable in the ADM formalism \cite{VWS01};
we recall that $\Gamma\equiv F'\equiv 2GM'$.
We note that \eqref{constraint1} is elliptic outside the horizon and
hyperbolic inside the horizon; this can be recognized from
\eqref{mathcalF}. 
In contrast to \cite{KMHV06,KMHSV07} and other papers, 
no additional factor ordering terms are taken into account here; this
will be crucial in obtaining our results. 

We now make the ansatz
\be
\lb{ansatz}
\Psi\left[\tau(r),R(r),\Gamma(r)\right]=
\exp\left(\frac{\I}{2l_{\rm P}^2}\int\D r\ \Gamma S(R,\tau)\right)\ ,
\ee
where $S(R,\tau)$ is a function to be determined recursively in the following
semiclassical approximation scheme. With the special factor ordering
chosen in \cite{KMHV06,KMHSV07}, this ansatz would lead to an {\em
  exact} solution of \eqref{constraint1} and \eqref{constraint2} with
semiclassical form. Here, instead,
we shall use this ansatz to solve \eqref{constraint1} in a
semiclassical approximation; the diffeomorphism constraint
\eqref{constraint2} is already solved identically with this ansatz.   

Inserting \eqref{ansatz} into \eqref{constraint1}, we arrive at
\bea
\lb{constraintnew}
& & \frac{\Gamma}{2}\left(\frac{\partial S}{\partial\tau}\right)^2
+\frac{\Gamma}{2}\cF\left(\frac{\partial S}{\partial R}\right)^2
-\frac{\Gamma}{2\cF}\nonumber\\ & & \;
 -\I l_{\rm P}^2\delta(0)\frac{\partial^2S}{\partial\tau^2}
-\I l_{\rm P}^2\delta(0)\cF\frac{\partial^2S}{\partial R^2}=0\ .
\eea
Here, ``$\delta(0)$'' indicates the presence of undefined expressions
arising from $\lim_{\bar{r}\to r}\delta(r-\bar{r})$, which require the
presence of a regularization scheme. (With the special factor ordering
chosen in \cite{KMHV06,KMHSV07}, the $\delta(0)$-terms are
automatically cancelled; this is why they were introduced there.) 

For the semiclassical approximation scheme, we make the ansatz
\be
\lb{ansatzS}
S=S_0+l_{\rm P}^2S_1+l_{\rm P}^4S_2+\ldots 
\ee
and compare consecutive orders in $l_{\rm P}^2$. (The general scheme
for such an approximation in quantum geometrodynamics is presented in
\cite{KS91,OUP}.)   
Since $\Gamma$ is dimensionless, we see from \eqref{ansatz} that $S$
has the dimension of a length ($L$), so that the dimension of $S_1$ is
$L^{-1}$, the dimension of $S_2$ is $L^{-3}$, and so on. 

Inserting \eqref{ansatzS} into \eqref{constraintnew} and comparing
different orders in $l_{\rm P}^2$, we obtain
\be
\lb{order0}
\cO\left(l_{\rm P}^0\right): \quad
\left(\frac{\partial S_0}{\partial\tau}\right)^2+\cF
\left(\frac{\partial S_0}{\partial R}\right)^2-\cF^{-1}=0\ ,
\ee
which, not surprisingly, is equivalent to the Hamilton--Jacobi
equation for the action
\be
\lb{action}
{\mathcal A}=\frac{1}{2G}\int\D r\ \Gamma S(R,\tau)\ .
\ee

 The next order yields
\bea
\lb{order2}
\cO\left(l_{\rm P}^2\right): \quad & & \Gamma\frac{\partial
  S_0}{\partial\tau} \frac{\partial
  S_1}{\partial\tau}+\Gamma\cF\frac{\partial S_0}{\partial R}\frac{\partial
  S_1}{\partial R}\nonumber\\ & & 
 -\I\delta(0)\frac{\partial^2S_0}{\partial\tau^2}
       -\I\delta(0)\cF\frac{\partial^2S_0}{\partial R^2}=0\ .
\eea
The following order then involves $S_2$,
\bea
\lb{order4}
\cO\left(l_{\rm P}^4\right): \quad & & \frac{\Gamma}{2}
\left[\left(\frac{\partial S_1}{\partial\tau}\right)^2+2\frac{\partial
S_0}{\partial\tau} \frac{\partial S_2}{\partial\tau}\right]\nonumber\\
& & \;
+\frac{\Gamma\cF}{2}\left[\left(\frac{\partial S_1}{\partial
      R}\right)^2 +2\frac{\partial
S_0}{\partial R} \frac{\partial S_2}{\partial R}\right]\nonumber\\
& & \; -\I\delta(0)\frac{\partial^2S_1}{\partial\tau^2}
-\I\delta(0)\frac{\partial^2S_1}{\partial R^2}=0\ .
\eea
The solution to the Hamilton--Jacobi equation \eqref{order0} is
\be
\lb{HJsolution}
S_0=\pm\left(a\tau+\int^R\D R\
  \frac{\sqrt{1-a^2\cF}}{\cF}\right)+\mathrm{constant} \ ,
\ee
where $a=1/\sqrt{1+2E}$. With the special factor ordering chosen in 
\cite{KMHV06,KMHSV07}, this would already be the exact solution to the
Wheeler--DeWitt equation (this, again, was part of the motivation to choose
that factor ordering in the first place). Here, however, the solution
\eqref{HJsolution} occurs at the highest order and must be used in the
next-order equations. 

The variable $a$ above is seen to be related to the (dimensionless)
energy $E$. This is similar to what happens for the tunneling mechanism
\cite{Paddy,BM08}.  
The corresponding ansatz there involves $\omega t$ in place of
$a\tau$, where $t$ is the Schwarzschild time and $\omega$ is
identified with the conserved quantity (in this case, the energy)
corresponding to the time-like Killing vector. 
We also note from (\ref{HJsolution}) since $a$ is
    dimensionless and $S_0$ has dimension of $L$, $\tau$ has also the
    dimension of $L$. This will be used later on. 

Observe that the solution \eqref{HJsolution} also holds in the
presence of a cosmological constant $\Lambda$, with $\cF$ then given
by $\cF=1-F/R-\Lambda R^2/3$ instead of \eqref{mathcalF}, see \cite{FGK09}
for the treatment of positive $\Lambda$ and \cite{VTS08} for the treatment of
negative $\Lambda$. In
that case, however, the integral can only be evaluated in terms of
elliptic functions. Furthermore, the constant in
\eqref{HJsolution} only enters the unknown total normalization
for $\Psi$ and will therefore be skipped in what follows.

Inserting now \eqref{HJsolution} into \eqref{order2}, we arrive at an
equation for $S_1$,
\be
\lb{S1equation}
a\Gamma\frac{\partial
  S_1}{\partial\tau}+\Gamma\sqrt{1-a^2\cF}\frac{\partial S_1}{\partial
  R}+\I\delta(0)\frac{F}{R^2}\frac{2-a^2\cF}{2\cF\sqrt{1-a^2\cF}} =0\ .
\ee
We solve this equation by the special ansatz 
\be
\lb{ansatzS1}
S_1=-C\tau -\I U_1(R)\ ,
\ee
where $C$ is a new variable with the dimension $L^{-2}$ because $S_1$
has dimension $L^{-1}$ and $\tau$ has dimension $L$; it will play 
a crucial role below. Since the only length scale in this model is
$GM$, we can set for later purpose
\be
\lb{alpha1}
C\equiv\frac{\alpha_1}{(GM)^2}\ ,
\ee
where $\alpha_1$ is a dimensionless constant. 

Inserting \eqref{ansatzS1} into \eqref{S1equation}, we get a
differential equation for $U_1(R)$,
\be
\lb{U1equation}
\frac{\D U_1}{\D R}=\frac{\I aC}{\sqrt{1-a^2\cF}}+
\delta(0)\frac{F}{\Gamma R^2}\frac{2-a^2\cF}{2\cF(1-a^2\cF)}\ .
\ee
We remark that in the marginal limit, $a\to 1$, this equation reads
\bdm
\frac{\D U_1}{\D R}=\I
C\sqrt{\frac{R}{F}}+\frac{\delta(0)}{2\Gamma(R-F)}
\left(1+\frac{F}{R}\right)  .
\edm
Integrating \eqref{U1equation}, we find apart from an irrelevant
constant the desired expression for $S_1$,
\bea
\lb{S1}
& & S_1(R,\tau) = -C\tau +aC\left[\frac{\sqrt{R([1-a^2]R+a^2F)}}{1-a^2}\right.
\nonumber\\ & & \; \left.-\frac{a^2F}{(1-a^2)^{3/2}}
\ln\left(\sqrt{(1-a^2)R}+\sqrt{(1-a^2)R+a^2F}\right)\right]
\nonumber\\
& & \; -\I\frac{\delta(0)}{2\Gamma}\left(2\ln(R-F)-\ln(R+a^2[F-R])-\ln
  R\right) \ .
\eea
We also give the special result for the marginal limit, $a\to 1$:
\be
\lb{S1marginal}
S_1(R,\tau)
=-C\tau+\frac{2CR^{3/2}}{3\sqrt{F}}-
\I\frac{\delta(0)}{2\Gamma}\left(2\ln(R-F)-\ln R\right) \ . 
\ee
In the next order, $\cO\left(l_{\rm P}^4\right)$, we have to insert
the solution \eqref{S1}, apart from (\ref{HJsolution}), into
\eqref{order4}. This yields a rather 
complicated equation for $S_2$. For example, in the special case $a=1$,
it is of the form 
\bea
& & \frac{\Gamma}{2}\left(C^2\pm 2\frac{\partial
    S_2}{\partial\tau}\right)
\nonumber\\ 
& & +\frac{\Gamma{\mathcal
    F}}{2}\left(\left[\frac{C}{\sqrt{1-{\mathcal F}}}-
\frac{\I\delta(0)}{2\Gamma}\frac{R+F}{R(R-F)}\right]^2\pm
2\frac{\sqrt{RF}}{R-F}\frac{\partial S_2}{\partial
  R}\right)\nonumber\\
& & -\I\delta(0){\mathcal
  F}\left(\frac{CF}{2R^2}\left[\frac{F}{R}\right]^{-3/2}
  -\frac{\I\delta(0)}{2\Gamma}
  \frac{(R-F)^2-2R^2}{R^2(R-F)^2}\right)\nonumber\\ & & \; =0\ .
\eea
In analogy to \eqref{ansatzS1}, one
could try to solve it with an ansatz of the form
\be
\lb{S2D}
S_2(R,\tau)=-D\tau-\I U_2(R)\ ,
\ee
where $D\equiv\alpha_2/(GM)^4$ has dimension $L^{-4}$ and
    involves another dimensionless constant $\alpha_2$. We shall
not, however, follow this here and restrict our 
attention only to the order $l_{\rm P}^2$.  

Collecting the solutions up to $\cO\left(l_{\rm P}^2\right)$, we can
write
\bea
\lb{S0S1}
& & S=S_0+l_{\rm P}^2S_1=
(\pm a-l_{\rm P}^2C)\tau\pm\nonumber\\ & & \;\;
\int^R\D R\ \frac{\sqrt{1-a^2\cF}}{\cF}\nonumber\\ & & 
+l_{\rm P}^2aC\tilde{G}(R)-\I l_{\rm P}^2\delta(0)\tilde{H}(R)\ ,
\eea
where
\bea
\lb{Gtilde}
\tilde{G}(R)&\equiv& \frac{\sqrt{R([1-a^2]R+a^2F)}}{1-a^2}
-\nonumber\\ & & \; \frac{a^2F}{(1-a^2)^{3/2}}
\ln\left(\sqrt{(1-a^2)R}\right.\nonumber\\ & & \; \left.
+\sqrt{(1-a^2)R+a^2F}\right)\ ,
\eea
and
\be
\lb{Htilde}
\tilde{H}(R)=\frac{1}{2\Gamma}\left(2\ln(R-F)-\ln(R+a^2[F-R])-\ln
  R\right)\ .
\ee
In analogy to earlier papers, cf. \cite{FGK09}, we define positive and
negative energy states according to 
the sign in front of the dust proper time variable $\tau$, with the
case of the minus sign corresponding to positive energy. Inserting
\eqref{S0S1} into the general ansatz \eqref{ansatz}, the
positive-energy solution reads
\bea
\lb{positive}
\Psi^+ &=&\exp\left(\frac{\I}{2l_{\rm P}^2}\int\D r\
  \Gamma\left[-a\tau-\int^R\D R\ \frac{\sqrt{1-a^2\cF}}{\cF}\right.\right.
\nonumber \\
& & \; \left. \left. -l_{\rm P}^2C\tau+l_{\rm P}^2aC\tilde{G}-\I l_{\rm
  P}^2\delta(0)\tilde{H}\right]\right)\ ,
\eea
while the negative-energy solution is given by
\bea
\lb{negative}
\Psi^-&=&\exp\left(\frac{\I}{2l_{\rm P}^2}\int\D r\
  \Gamma\left[a\tau+\int^R\D R\ \frac{\sqrt{1-a^2\cF}}{\cF}\right.\right.
\nonumber\\
& & \; \left.\left. -l_{\rm P}^2C\tau+l_{\rm P}^2aC\tilde{G}-\I l_{\rm
  P}^2\delta(0)\tilde{H}\right]\right)\ .
\eea
In order to calculate the Hawking radiation, we shall evaluate the
overlap between the ``outgoing dust state with negative energy''  
$\Psi^-_{\rm e}$ (where the index ``e'' refers to ``expanding'' cloud) and
the ``ingoing dust state with positive energy'' $\Psi^+_{\rm c}$
(where the index c refers to ``collapsing'' cloud). 
 Since the interpretation of these states is made with respect to an 
observer in the asymptotic regime using the Killing time $T$, we have to 
substitute the dust time $\tau$ by $T$ \cite{FGK09}. For the outgoing
case, we have the relation
\be 
\lb{killingplus} 
T=a \tau + \int \D R \ \frac{\sqrt{1-\mathcal{F} a^2}}{\mathcal{F}}\ , 
\ee 
while in the ingoing case we have
\be 
\lb{killingminus} 
T=a \tau - \int \D R \ \frac{\sqrt{1-\mathcal{F} a^2}}{\mathcal{F}}\ . 
\ee 
For the concrete calculation we shall write the full states as a 
product of single-shell states where the radial variable $r$ is 
assumed to consist of discrete points separated by a distance 
$\sigma$. (The continuum limit is obtained for $\sigma\to 0$.) As in 
\cite{KMHSV07}, the Bogolyubov coefficient $\beta$ is calculated for 
each shell separately. In the discrete case, we replace $\Gamma$ by 
the dimensionless variable $2\omega$ and indicate the dependence on 
$\omega$ by an index. (The factor $2$ is motivated by 
the fact that $\Gamma=2GM'$.) 
We omit the shell index and write the 
corresponding wave functions as $\psi_{\omega}(T,R)$.  
We then define $\beta$ to read 
\be 
\lb{beta} 
\beta_{\omega}\equiv \int_F^{\infty}\D R\ 
\sqrt{g_{RR}}\Psi^{-*}_{{\rm e}\omega} \Psi^+_{{\rm c}\omega}\ ,
\ee 
where $g_{RR}$ is the $RR$-component of the DeWitt metric, as it can 
be read off \eqref{constraint1} where the inverse of the DeWitt metric is 
$l_{\rm P}^{-4}$ times the prefactor of 
the term $\delta^2\Psi/\delta R^2$. We thus have $g_{RR}={\mathcal
  F}^{-1}$; performing then the required coordinate transformation
from the variables $(\tau,R)$ to $(T,R)$ gives the result
$\sqrt{g_{RR}}=(a{\mathcal F})^{-1}$ which has to be used in the
calculation of the Bogolyubov coefficient $\beta$.  

Inserting now $\Psi^{-*}_{{\rm e}\omega}$ and $\Psi^+_{{\rm c}\omega}$
into \eqref{beta}, we get
\bea
\lb{betaintegral}
\beta_{\omega}&=&\int_F^{\infty}\D R\
(a\cF)^{-1}\exp\left(-\frac{2\I\omega\sigma}{l_{\rm
      P}^2}\left[T+\right.\right.\nonumber\\ & & \left.
\int^R\D R\ \frac{\sqrt{1-a^2\cF}}{\cF}\right]
\nonumber\\
 & &  \left. -2\I\omega\sigma\frac{C}{a}\int^R\D R\ 
\frac{\sqrt{1-a^2\cF}}{\cF}+2\omega\sigma\delta(0)\tilde{H}\right)\ .
\eea
We note that this expression is independent of $\tilde{G}$.

In the following, we shall employ a ``DeWitt regularization'' and set
$\delta(0)=0$. Whether this can consistently be done at the most
fundamental level is, however, not clear at this stage; here, it is
merely used as a formal recipe.  
Recalling that $(a\cF)^{-1}=R/a(R-F)$, we then get
\bea
\beta_{\omega}&=&a^{-1}\exp\left(-\frac{2\I\omega\sigma T}{l_{\rm P}^2}\right)
\int_F^{\infty}\D R\ \frac{R}{R-F}\times\nonumber\\
& & \; \exp\left(-\frac{2\I\omega\sigma}{l_{\rm P}^2}\left[1+
l_{\rm P}^2\frac{C}{a}\right]\frac{\sqrt{1-a^2\cF}}{\cF}\right)\ .
\eea
As in \cite{KMHSV07}, we introduce the dimensionless integration
variable
\bdm
s=\sqrt{\frac{R}{F}}-1
\edm
and get
{\small
\bea
\lb{betaintegral2}
& & \beta_{\omega}=2Fa^{-1}\exp\left(-\frac{2\I\omega\sigma T}
{l_{\rm P}^2}\right)\int_0^{\infty}\D s\ \frac{(1+s)^3}{s^2+2s}\times
\nonumber\\
& &  \exp\left(-\frac{4\I\omega\sigma F}{l_{\rm
      P}^2}\left[1+
l_{\rm P}^2\frac{C}{a}\right]\int^s\D s\right. \nonumber\\ & &\;\;\; \left.
\frac{(s+1)^2}{s^2+2s}\sqrt{(1+s)^2-a^2(s^2+2s)}\right).
\eea
}
Up to higher orders of the Planck length squared in the exponent, this
is so far an exact expression. As in \cite{KMHSV07}, we now assume
that the $s$-integral from zero to infinity is dominated by its
contribution near $s=0$, that is, near the horizon; this is also the
assumption in the standard derivation of the Hawking effect \cite{Hawking1}.
Using
therefore in \eqref{betaintegral2} the approximation
\bea
\lb{approximation}
& & \frac{(s+1)^2}{s^2+2s}\sqrt{(1+s)^2-a^2(s^2+2s)}=\nonumber\\
& & \; \frac{1}{2s}
\left(1+\left[\frac{5}{2}-a^2\right]s+{\mathcal O}(s^2)\right)\ , 
\eea
we get
\bea
\lb{approx}
& & \beta_{\omega}\approx Fa^{-1}\exp\left(-\frac{2\I\omega\sigma T}
{l_{\rm P}^2}\right)\int_0^{\infty}\D s\ s^{-1-\frac{2\I\omega\sigma
  F}{l_{\rm P}^2}(1+l_{\rm P}^2\frac{C}{a})}\times
\nonumber\\
& & \; \exp\left(-\frac{2\I\omega\sigma F}{l_{\rm P}^2}
\left[1+l_{\rm P}^2\frac{C}{a}\right]\left[\frac{5}{2}-a^2\right]s\right)\ .
\eea
To evaluate this integral, we use the formula \cite{gradshteyn}
\begin{displaymath} 
\int_0^{\infty}\md x\ x^{\nu-1}e^{-(p+\I q)x}=\Gamma(\nu)(p^2+q^2)^{-\nu/2} 
  e^{-\I\nu\mathrm{arctan}(q/p)} \ ,
\end{displaymath} 
which is, in particular, applicable to the case $p=0$ and $0<{\rm
  Re}\ \nu<1$.
(We insert a small positive value for ${\rm Re}\nu$, which we let go
to zero after the integration.) 
Using, moreover,
\bdm 
\Gamma\left(-\I u\right)\Gamma\left(\I u\right)=\frac{\pi}{u\sinh \pi
  u}
\edm 
(with real $u$), we get 
\be
\lb{betasquared}
\vert\beta_{\omega}\vert^2 \approx\frac{2\pi F^2}{a^2y}\frac{1}{\E^{2\pi y}-1}
\ee
with
\be
y=\frac{2\omega\sigma F}{l_{\rm P}^2}\left(1+l_{\rm
    P}^2\frac{C}{a}\right)\ .
\ee
Substituting $\sigma\omega$ by $G\Delta\epsilon$,\footnote{Recall that
  $\omega$ is the discretized version of $\Gamma/2=GM'$, so
  $\sigma\omega$ corresponds to $G\Delta M\equiv G\Delta\epsilon$.} 
where 
$\Delta\epsilon$ is the energy of a shell, 
and introducing the physical frequency $\Omega=\Delta\epsilon/\hbar$,
we arrive at the final result 
\be
\lb{betafinal}
\vert\beta_{\Omega}\vert^2=\frac{2\pi GM}{\Omega a^2\left(1+l_{\rm
P}^2\frac{C}{a}\right)}\frac{1}{\exp\left(\frac{\hbar\Omega}{k_{\rm B}T_{\rm
H}}\right)-1} 
\ee
with the quantum-gravity corrected Hawking temperature
\be
\lb{TH}
k_{\rm B}T_{\rm H}=\frac{\hbar}{8\pi GM\left(1+l_{\rm
      P}^2\frac{C}{a}\right)} \ .
\ee

A number of remarks are in order.
First, the meaning of the prefactor in the expression for
$\vert\beta_{\Omega}\vert^2$ (which depends mildly on $\Omega$) is
unclear. It is certainly connected with the greybody factors, but
without a clear-cut normalization of the quantum states, its
interpretation remains incomplete. Secondly, different from the
earlier papers, we have calculated the overlap of the quantum states
in \eqref{beta} for coinciding frequencies $\omega$ only. The reason
is that an (approximate) thermal spectrum only occurs in that
case. Unlike the highest order $l_{\rm P}^0$, taking here 
also into account two frequencies
$\omega$ and $\omega'$ corresponding to two different shells,
the integration over $\omega'$ would spoil the
thermality. The results \eqref{betafinal} and \eqref{TH} thus remain
valid only as far as the interaction between different shells is
subdominant. Thirdly, when taking the next order in the Planck-mass
expansion into account, we
expect that the term $1+l_{\rm P}^2C/a$ in the denominator of
\eqref{TH} is augmented by a term proportional to $l_{\rm P}^4D/a$,
where $D$ occurs in \eqref{S2D}.

The form of the temperature given in \eqref{TH} for $a=1$ was obtained earlier
for the case of the Schwarzschild black hole in \cite{BM08}. It was
calculated there by using the quantum tunneling method beyond the
semiclassical approximation.

We emphasize that in the previous papers \cite{KMHSV07,FGK09} no such
quantum gravitational correction to the Hawking temperature has been
found, since calculations have led to an exact solution with
semiclassical form.

Substituting now (\ref{alpha1}) in (\ref{S1}) we obtain the expression
for $S_1$ as, 
\bea
\lb{S11}
& & S_1(R,\tau) = -\frac{\alpha_1}{(GM)^2}\tau +\nonumber\\ & & 
a\frac{\alpha_1}{(GM)^2}\left[\frac{\sqrt{R([1-a^2]R+a^2F)}}{1-a^2}\right. 
-\nonumber\\ & & \left.\frac{a^2F}{(1-a^2)^{3/2}}
\ln\left(\sqrt{(1-a^2)R}+\sqrt{(1-a^2)R+a^2F}\right)\right]
\nonumber\\
& & -\I\frac{\delta(0)}{2\Gamma}\left(2\ln(R-F)-\ln(R+a^2[F-R])-\ln
  R\right) 
\eea

We discuss in the following a method developed in \cite{BM09} to find
the value of the 
dimensionless constant $\alpha_1$.  
Consider for that purpose a constant scale transformation of the
coefficients of the 
metric (\ref{1.56}), given by (cf. also \cite{Hawking2})
\begin{eqnarray}
{\bar{g}}_{\mu\nu}=kg_{\mu\nu}\ .
\label{1.58}
\end{eqnarray}
Under this transformation we have from (\ref{1.56}),
\begin{eqnarray}
\bar{R}=k^{\frac{1}{2}}R\ .
\label{1.59}
\end{eqnarray}
For the Einstein equations to remain invariant under this scale
transformation, $F$ should according to \eqref{ltb-eg2} transform as
\begin{eqnarray}
\bar{F}=k^{\frac{1}{2}}F\ ,
\label{1.60}
\end{eqnarray}
and $\tau$ should transform as 
\be
\label{rev4}
\bar{\tau}=k^{\frac{1}{2}}\tau \ .
\ee
Therefore, (\ref{mathcalF}) yields
\begin{eqnarray}
\bar{\cal{F}}={\cal{F}}\ ,
\label{rev1}
\end{eqnarray}
and since $F=2GM$, $GM$ transforms as
\begin{eqnarray}
\overline{(GM)}=k^{\frac{1}{2}}(GM)\ .
\label{1.61}
\end{eqnarray}
 For the Wheeler--DeWitt equation (\ref{constraint1}) to be invariant
 under this scale transformation, we must have in addition the following
 transformations: 
\begin{eqnarray}
&&\bar{\Psi}=k\Psi\ ,
\label{rev2} 
\\
&&\bar{\Gamma}=k^{-\frac{1}{2}}\Gamma\ ,
\label{rev3}
\end{eqnarray}
where (\ref{1.59}) and (\ref{rev1}) have been used.

Now from the expression for $\Psi$, Eq. (\ref{ansatz}), we have
\begin{eqnarray}
\frac{\hbar}{\I}\frac{\delta\Psi (\Gamma,R,\tau)}{\delta
  X^{\mu}(r)}=\frac{\Gamma (r)}{2G}\frac{\partial S(R,\tau)}{\partial
  X^{\mu}(r)}\Psi(\Gamma,R,\tau)\ , 
\label{rev5}
\end{eqnarray}
where $\mu=0,1$ with $X^0=\tau$ and $X^1=R$.
Because we have in \eqref{rev5} a functional derivative on the
left-hand side and an ordinary 
  derivative multiplied by $\Gamma(r)$ on the right-hand side,
  $S$ is not the  
  action. In fact, as we have seen, it is the quantity \eqref{action}
  which has the correct physical dimension mass times length and which
  is equal to the action of this model. The first-order action is
  therefore given by
\be
\lb{firstorder}
{\mathcal A}_1=\frac{l_{\rm P}^2}{2G}\int\D r\ \Gamma S_1(R,\tau)\ .
\ee
We have seen that the relevant part for the recovery of Hawking
radiation was the first, purely $\tau$-dependent, term in
\eqref{S11}. Because this part does not contain the gravitational
variables, it can for our purpose be considered as the matter (dust)
action which we shall call ${\mathcal A}_1^{\rm m}$. We thus have
\be
{\mathcal A}_1^{\rm m}= -\frac{l_{\rm P}^2}{2G}\int\D r\ \Gamma
\frac{\alpha_1\tau}{(GM)^2} \ .
\ee 
Hence, under the transformations \eqref{rev4}, \eqref{1.61}, and
\eqref{rev3}, ${\mathcal A}_1^{\rm m}$ transforms as
\be
\bar{{\mathcal A}_1^{\rm m}}=-\frac{l_{\rm P}^2}{2G}\int\D r\
\bar{\Gamma}\frac{\alpha_1\bar{\tau}}{(\overline{GM})^2}
=k^{-1}{\mathcal A}_1^{\rm m} \simeq(1-\delta k){}{\cal{A}}_1^{\rm m} \ ,
\ee
where we have assumed that $\alpha_1$ and $r$ do not scale, and $k\simeq
1+\delta k$.
Hence,
\begin{eqnarray}
\delta {\cal{A}}_1^{\rm m}= \bar{{\cal{A}}}_1^{\rm m}-{\cal{A}}_1^{\rm
  m}=-{\cal{A}}_1^{\rm m}\delta k\ .
\label{1.65}
\end{eqnarray}
Now using the definition of the energy--momentum tensor,
\begin{eqnarray}
& & \int d^4x\sqrt{-g}T^\mu_\mu=\frac{2\delta {\cal{A}}_1^{\rm m}}{\delta
  k}=-2{\cal{A}}_1^{\rm m}\nonumber\\ & & \;
=\hbar\int\D r\ \Gamma\frac{\alpha_1\tau}{(GM)^2}
\longrightarrow 2\omega\sigma \frac{\hbar\alpha_1\tau}{(GM)^2}\ ,
\label{1.66}
\end{eqnarray}
where in the last step, we pass to the discretized version by
replacing $\Gamma$ by the dimensionless variable $2\omega$ as before.

Therefore, by a simple interposition of (\ref{1.66}), we obtain, 
\begin{eqnarray}
\alpha_1=\frac{(GM)^2}{2\hbar\omega\tau\sigma}\int \D^4x\sqrt{-g}T^\mu_\mu\ .
\label{1.67}
\end{eqnarray}

    Next, our task is to find the value of $\tau$. For the {\it
      contracting cloud} case, $\tau$ is given by the relation
    (\ref{killingminus}), which can be written in a rearranged form:
\begin{eqnarray}
\tau = \frac{T}{a} + \frac{1}{a} \int \D R\
\frac{\sqrt{1-{\cal{F}}a^2}}{\cal{F}}\ . 
\label{rearrange}
\end{eqnarray}
Now in the calculation of $\Big|\beta_\omega\Big|^2$ (see equations
(\ref{approx}) and (\ref{betasquared})), which eventually gives the
flux for the model, the first term in (\ref{rearrange}) is
inconsequential, since it occurs as a phase factor ($\sim \E^{\I\alpha
  T}$) which yields unity on taking the modulus. 
On the other hand,  as we have seen, the terms which
    contribute to the flux come from the integration term. Therefore, for
    the present calculation, considering only the last term of
    (\ref{rearrange})  
and using the fact that ${\cal{F}}=1-\frac{F}{R}$ and then substituting
$s=\sqrt{\frac{R}{F}}-1$, we obtain, 
\begin{eqnarray}
\tau=2F\int \D s\ \frac{(s+1)^2}{s^2+2s}\sqrt{(s+1)^2 -
  a^2(s^2+2s)}\ .
\label{1.68n1}
\end{eqnarray}
If this integral is supposed to run from $0$ to $\infty$ (since $R$
runs from $F$ to $\infty$), it is certainly divergent. In order to
make contact with the expressions in earlier papers, we make the
following heuristic considerations, which can be viewed as a
regularization prescription. Restricting attention to the relevant
regime near $s=0$, 
use of the approximation (\ref{approximation}) in the above yields,
\begin{eqnarray}
\tau \simeq 2F\int \D s\
\frac{1}{2s}\Big[1+\left(\frac{5}{2}-a^2\right)s\Big]
=F\ln s+{\mathcal O}(s).
\label{tau1}
\end{eqnarray} 
Interpreting $s$ as a complex variable and recalling $\ln s=\ln\vert
s\vert+\I{\rm arg}s$, we can define a ``Euclidean time'' $\tau_{\rm
  E}$ by
\begin{eqnarray} 
\tau_{\rm E}=\I\tau=-\pi F=-2\pi(GM)\ .
\label{1.68n2}
\end{eqnarray}

Finally, substituting this value for $\tau_{\rm E}$ in 
the Euclideanized version of (\ref{1.67}), we obtain
our cherished expression, 
\be
\alpha_1=-\frac{GM}{4\pi\hbar\omega\sigma}\int \D^4x_{\rm
  E}\sqrt{-g}T^\mu_\mu\ .
\label{1.69}
\ee
This shows that
$\alpha_1$ is related to the ``one-loop trace anomaly'' of the
energy--momentum tensor. A similar result was obtained earlier in
\cite{BM09,CL,DF,BM,RMO} for the eternal black hole case.

     To conclude, we mention that a quantum gravitational correction to the Hawking temperature from the LTB model was established through a semiclassical approximation scheme employed in \cite{BM08}. Here, no special factor ordering \cite{KMHV06,KMHSV07} was chosen, which was a crucial step to obtain such a correction. Instead, we considered all the terms in the expansion for $S$ (\ref{ansatzS}). This led to several equations corresponding to different orders in $l_P^2$. In this paper, only terms upto ${\cal{O}}(l_P^2)$ were considered. The equations were solved by a special ansatz. After getting the solutions for the states upto ${\cal{O}}(l_P^2)$, the ``De Witt regularization'' was employed. This regularization enforced $\delta(0) = 0$. The explicit calculation of the Bogolyubov coefficient near the horizon led to the emission spectrum. The corrected Hawking temperature was then automatically identified. It contained an unknown variable ``$C$''. Dimensional arguments then helped us to fix ``$C$'', apart from a dimensionless constant.

     The last part of the paper was actually devoted to fix the dimensionless constant appearing in ``$C$''. It was done by a constant scale transformation of the metric coefficients (\ref{1.56}). A detailed analysis showed that it was related to the one loop trace anomaly of the energy - momentum tensor for the dust (matter).

       It must be emphasized that a similar result was obtained by Hawking \cite{Hawking2} for an eternal black hole space-time by taking into account the one loop correction to the partition function due to the fluctuations of the scalar fields on the black hole space-time. Exactly the same result was also derived later on by different methods \cite{BM09,CL,DF,BM,RMO}. Here our analysis was done in the spirit of the quantum tunneling method employing the WKB approximation \cite{BM08,BM09,BM,RMO}. Indeed the special ansatz (\ref{ansatzS1}) used here closely resembles the Hamilton - Jacobi splitting of the one particle action $S(t,r) = \omega t + {\tilde{S}}(r)$. Such a similarity of our result with Hawking's finding \cite{Hawking2} may be due to the equivalence of the path integral with the WKB ansatz upto ${\cal{O}}(l_P^2)$, a result that has been established earlier in quite general terms \cite{Morette}. That this connection also holds in the black hole context is a new observation.

\section*{Acknowledgements} We thank Max D\"orner for helpful
discussions and critical comments. One of the authors (RB) would like
to thank the Alexander von Humboldt Foundation for providing financial
support and members of the Gravity and Relativity group, University of
K$\ddot{\textrm{o}}$ln, where a major part of this work was done, for
their gracious hospitality.


 
\end{document}